\documentstyle[epsf,preprint,prl,aps]{revtex}
\tightenlines
\newcommand{\floor}{{\rm f\/loor}}  
\newcommand{\mod}{{\rm mod}}
\newcommand{\tprime}{ {t^\prime} }
\newcommand{\jprime}{ {j^\prime} }

\newcommand{\beq}{\begin{equation}}
\newcommand{\eeq}{\end{equation}}

\newcommand{\der}[2]{\frac{d #1}{d #2}}

\newcommand{\half}{\frac{1}{2}}

\newcommand{\beqar}{\begin{eqnarray}}
\newcommand{\eeqar}{\end{eqnarray}}

\newcommand{\ep}{\epsilon}

\newcommand{\rar}{\rightarrow}

\def\k{{\mbox{\boldmath $ k$}}}
\title{Noise Stabilization of Self-Organized Memories}
\author{
M.L. Povinelli\thanks{Present address:
Physics Dept., MIT, Cambridge,
MA 02139}, S.N. Coppersmith, L.P. Kadanoff,
S.R. Nagel, and S.C. Venkataramani\thanks{Present address:
Math Dept., University of Chicago, 1100 E.
58th St., Chicago, IL 60637}}
\address{The James Franck Institute, The University of Chicago,
5640 Ellis Avenue, Chicago, IL  60637}

\begin{document}
\maketitle
\centerline{\today}

\begin{abstract}
We investigate a nonlinear dynamical system which
``remembers'' preselected values of a system parameter.
The deterministic version of the
system can encode many parameter values during a transient
period, but in the limit of long times, almost all of them
are forgotten.
Here we show that a certain type of stochastic
noise can stabilize multiple memories, enabling many parameter
values to be encoded permanently.
We present analytic results that provide insight
both into the memory formation and into the
noise-induced memory stabilization.
The relevance of our results to
experiments on the charge-density wave material $NbSe_3$
is discussed.
\end{abstract}

\section{Introduction}
This paper concerns a nonlinear dynamical system with
many degrees of freedom which organizes to store memories,
in that a configuration-dependent quantity is driven to
take on preselected values.
In Ref.~\cite{coppersmith97} it is shown that
in the absence of noise, the system encodes many memories
during a transient period, but in the limit of long times
retains no more than two of them.
Thus, the purely deterministic system ``learns,'' and then it
``forgets.''

We examine the effects of adding noise to this system
and demonstrate that certain types of noise can stabilize
multiple memories so that they are remembered permanently.
This noise stabilization is possible
because the memory formation mechanism is fundamentally
local, whereas forgetting is governed by the large-scale behavior
of the system.
Thus, it is possible for certain types of stochastic noise
to modify the behavior at long wavelengths
without destroying the local nonlinear dynamics which give
rise to memory creation.

We argue that the type of noise that we have found
to stabilize multiple memories is likely
to be present in some experiments on
charge-density wave (CDW) conductors such as $NbSe_3$.
Thus, our results could explain the experimental
observation of multiple apparently permanent memories
encoded in individual samples %that is
reported in Ref.~\cite{coppersmith97}.

Our analytic investigations of the behavior
of this system both with and without noise
show that insight into the mechanisms underlying
memory formation as well as noise stabilization can be obtained
by averaging the dynamical equations over intermediate time periods.
We determine analytically
the dependence of the memory values on the noise parameters
in the limit when a certain parameter $k$ tends to zero.
The large-scale behavior of the system follows
closely that of a linear diffusion equation; we present
analytic bounds on the differences between the evolution
of the nonlinear equations and that of the linearized
system that are uniform in
time and logarithmic in the system size.
Some of the analytic results for the system without noise
were asserted but not justified in Ref.~\cite{coppersmith97}.

The paper is organized as follows.  Section \ref{model}
briefly reviews the deterministic version of the model.
Sections \ref{simulations:methods} and \ref{simulations:results}
present our numerical work demonstrating
that noise can stabilize multiple memories.
Section \ref{analytics} presents our analytic work
which enables us to understand why noise can keep
memories from being forgotten and 
also presents an averaging procedure which allows us
to obtain analytic insight into the transient memories
present in the map without noise in a certain limit.
Section \ref{discussion} discusses our main results and
possible relevance to CDW experiments.
Appendix A demonstrates the uniqueness of the limit obtained
by the averaging procedure of section \ref{analytics} and also
discusses explicitly this limit for the case of
multiple memories.
Appendix B compares the time evolution of the full nonlinear
system with the time evolution of a linearized model and
shows that the linearized equations reproduce accurately
aspects of the evolution on large scales
(though not the memory formation itself).

\section{The model}
\label{model}
First we present the model with no noise, which is the
system of coupled maps studied in Ref.~\cite{coppersmith97}:
\begin{equation}
x_j(t+1) \ =\  x_j(t) + {\floor} \left [
k\sum_{i~(nn)}(x_i(t)-x_j(t))\ - A(t) \right ] .
\label{eq:xeq}
\end{equation}
Here, $i,j$ are the site indices, the sum is over nearest neighbors,
$t $ is a discrete time index,
and ${\floor} [z]$ is the largest integer less
than or equal to $z$.
This system of maps
can be derived from continuous-time differential equations describing
the purely dissipative evolution of the
positions $x_j$ of $N$ particles in a deep periodic potential, with nearest
neighbor particles connected by springs of spring constant $k \ll 1$
(see inset, figure 1),
in the presence of force impulses
$(-A(t)+\frac{1}{2})$\cite{coppersmith87b,ballfootnote,floria96}.
These equations describe the dynamics of sliding
charge-density waves (CDW's),\cite{coppersmith87b,cdwrefs,coppersmith87a},
and are closely related to models of a variety of dynamical
systems\cite{othermodels}.
In this paper we will consider explicitly only one-dimensional systems
of $N$ degrees of freedom with
one free and one fixed end, $x_0(t)=0$ and $x_{N+1}(t)=x_N(t)$,
starting from the initial condition $x_j(t=0)=0$.
However, it is straightforward to generalize
almost all the results to a variety of different boundary conditions
and to more dimensions.

The memory formation that occurs as these maps evolve is manifest
in the discrete curvature variables\cite{tang87}
\begin{equation}
\label{eq:defcurv}
c_j(t) \ =\ k\sum_{i~(nn)}(x_i(t)-x_j(t))~.
\end{equation}
It will be useful to write the dynamical equations
in terms of the curvature variables $c_j(t)$ rather than the
particle positions $x_j(t)$.
The equations of motion for the $c$'s are:
\begin{eqnarray}
\lefteqn{c_j(t +1) - c_j(t) = } \nonumber \\*
& & k \left \{
{\floor} \left [ c_{j+1}(t)  - A(t) \right ]
- 2 ~{\floor} \left [ c_{j}(t)  - A(t) \right ]
+ {\floor} \left [ c_{j-1}(t)  - A(t) \right ]
\right \} \ ;
\label{eq:curveq}
\end{eqnarray}
the fixed chain boundary conditions are\cite{bc_note}
\begin{eqnarray}
c_0(t)=A(t)~, \\
c_{N+1}(t)=c_N(t) ;
\label{eq:cboundary}
\end{eqnarray}
and the initial conditions are
\begin{equation}
c_i(t=0)=0 ~,~\ \ \ \mbox{$i \ne 0$}.
\label{eq:cinitial}
\end{equation}

Figure 1 shows for these initial conditions
the curvature variables $c_j(t)$ versus time $t $
for a five-particle chain when $A(t)$ is cycled
sequentially through five different values.
Memory encoding is manifest by the tendencies of $c$'s
to take on values where
${\rm frac}(c)= {\rm frac}(A(t))$, where
${\rm frac}(z) = z -\floor(z)$.
That the curvature variables take on values whose
fractional part is equal to the fractional part of the force impulse
in the maps implies
that for the balls and springs, just at the end of each force pulse
a significant fraction of the 
balls are found near the tops of their potential
wells\cite{oldcdwrefs}.
The system ``memorizes'' the force values and adjusts itself
so that the balls are at the well tops just as the pulses end.

As seen in Ref.~\cite{coppersmith97} and here in figure 1, if
a repeating sequence of pulses of different lengths is applied,
then for a while all the values are encoded.
However, at long times
the system eventually reaches a fixed point where it stops evolving.
At the fixed point, the curvatures are all the same;
only one memory is remembered.
We have found that using periodic, free, and fixed boundary conditions
for the $x$'s, changing the initial conditions, and incorporating
quenched (time-independent) randomness to the model do not increase
the number of memories retained at the fixed point beyond two.

Here we investigate how this system can be modified so that it can
remember more memories permanently.
We show that certain types of stochastic noise which
are likely to be present in some CDW experiments can do this
and study both numerically and analytically the reasons for
the multiple-memory stabilization.
We show that the stabilization of many memories arises
because the noise contains a deterministic component which
causes the curvature variables to
sustain a large scale spatial variation even in the limit of
infinite time.
The purely stochastic elements of the noise act to destabilize
the memories; we will see that these destabilizing effects vanish
in the limit $k \rightarrow 0$.

The following two sections presents our numerical investigations
of the model with noise.

\section{Technique: Adding Noise}
\label{simulations:methods}
Noise terms can be added to Eq.~(\ref{eq:xeq}) in a variety of ways.
Noise which is uncorrelated in both space and time, or
uncorrelated in time but spatially uniform
(e.g., fluctuations in the pulse amplitudes)
does not lead to memory stabilization.
However, we have identified a type of noise which is physically
plausible that stabilizes multiple memories.

The memory-stabilizing noise we study here
is defined by modifying Eq.~(\ref{eq:xeq}) as follows.
Every $\tau$ time steps, an index $j_D$  with $1 \le j_D \le N$
is selected, and for all $j>j_D$, the positions of
balls $j_D$ through $N$ are shifted by
a fixed integer $X$\cite{signfootnote}
\begin{equation}
\label{eq:noisemap}
x_j(t+1)=\left\{ \begin{array}{ll}
	x_j(t)+\floor[c_j(t)-A(t)]+X & \mbox{if $j \ge j_D$}  \\
	x_j(t)+\floor[c_j(t)-A(t)] & \mbox{otherwise.}
	\end{array}
\right.
\end{equation}
The relative positions of all the balls are unchanged
except for the distance between $x_{j_D}$ and $x_{j_{D}-1}$,
so the disruption is local.

Equivalently, one can write the map with noise in
terms of the curvature variables, as
\begin{eqnarray}
\label{eq:curvnoisemap}
c_j(t+1)-c_j(t) & = &
        k \left \{ \floor[c_{j+1}(t)-A(t)] 
	- 2 ~\floor[c_j(t)-A(t)] \right . \nonumber \\*
	& + & \left . \floor[c_{j-1}(t)-A(t)]  \right \}  +
	kX\delta_{t(\mod~\tau),0} ~(\delta_{j,j_D-1}-\delta_{j,j_D})~,
\end{eqnarray}
where the Kronecker delta $\delta_{i,j}$ is unity if $i$ and $j$
are identical and zero otherwise.
In either formulation, the noise does not affect the
boundary conditions.

This type of noise models the physical process of
breaking the spring connecting balls $j_D$ and $j_{D}-1$ and then
subsequently reconnecting them with a spring of
longer unstretched length\cite{springfootnote}.
This choice of noise is motivated by phase slip processes known to occur
in CDW materials\cite{phasesliprefs}.
When the model is applied to CDW's, the variable $x_j$ in
Eq.~(\ref{eq:noisemap}) is interpreted as
the phase of the charge density wave at the $j^{th}$ impurity site
in the sample, measured relative to an undistorted
configuration\cite{pietronero83,fisher85,littlewood86}.
If a phase slip causes an extra wavelength
of the CDW to be inserted between two impurities, then the ``unstretched''
phase difference between two impurity sites increases.

\section{Numerical Results}
\label{simulations:results}
\begin{figure}
\centerline{\epsfxsize=\hsize \epsfbox{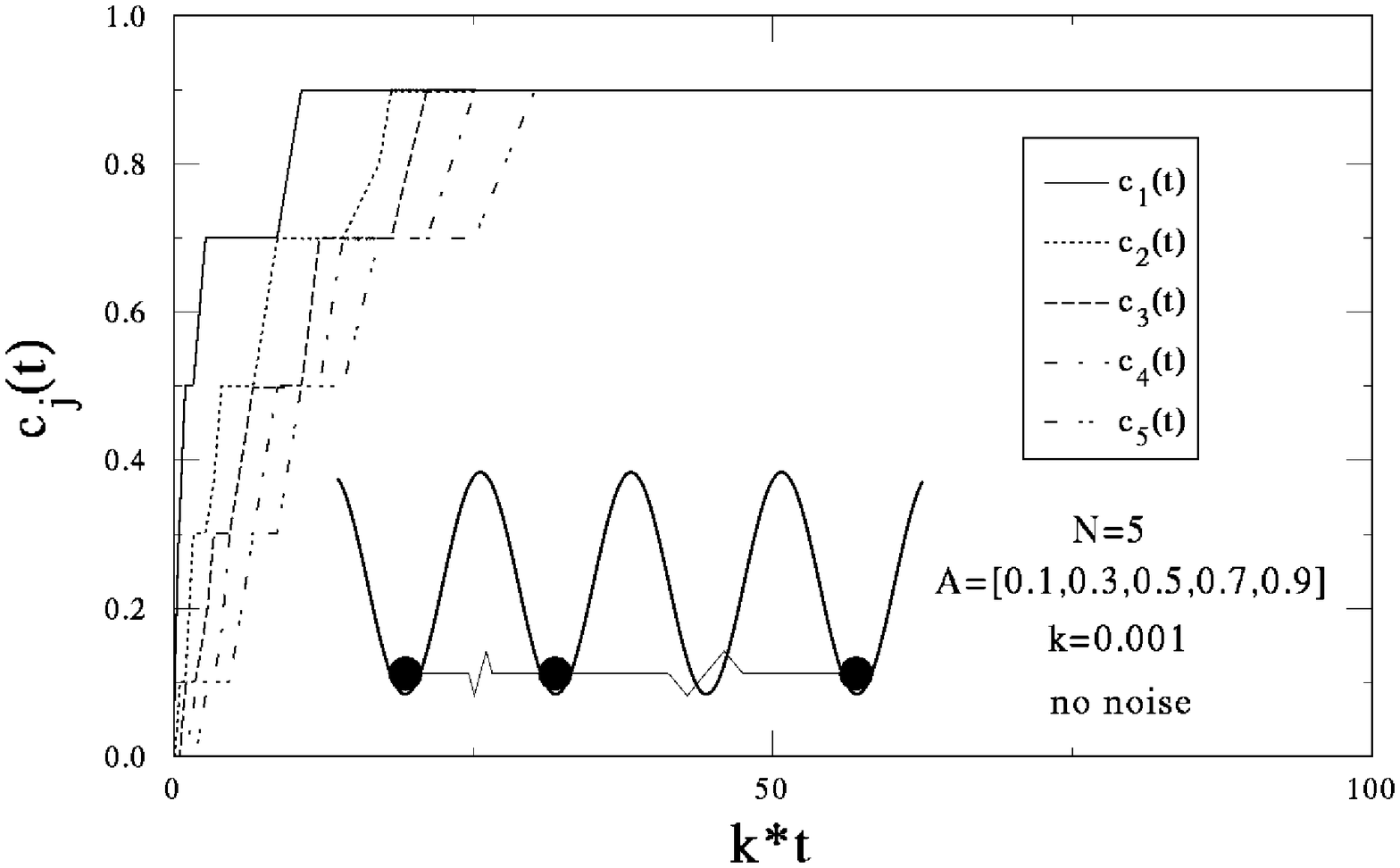}}
\vspace{1cm}
\caption{Plot of curvatures $c_j(t) = k(x_{j+1}(t) - 2x_j(t)+x_{j-1}(t))$
versus scaled time variable $k*t$
for Eqs.~(\protect\ref{eq:xeq}) with no noise and boundary conditions
$x_{0}(t)=0$, $x_{N+1}(t)=x_N(t)$,
starting from the initial condition $x_j(t=0)=0$ for $j=1,\ldots,N$.
System parameters are given on the plot.
The horizontal regions in the graphs
occur when the fractional part of one of the curvature values
equals the fractional part of one of the values of $A$.
Notice that while the balls spend some time on each of the memory values,
all the curvature values eventually
end up at the single memory value $0.9$.
Inset: sketch of balls and springs in periodic potential, a physical
realization of Eqs.~(\protect\ref{eq:xeq}).}
\label{fig:nonoise}
\end{figure}

\begin{figure}
\centerline{\epsfxsize=.8\hsize \epsfbox{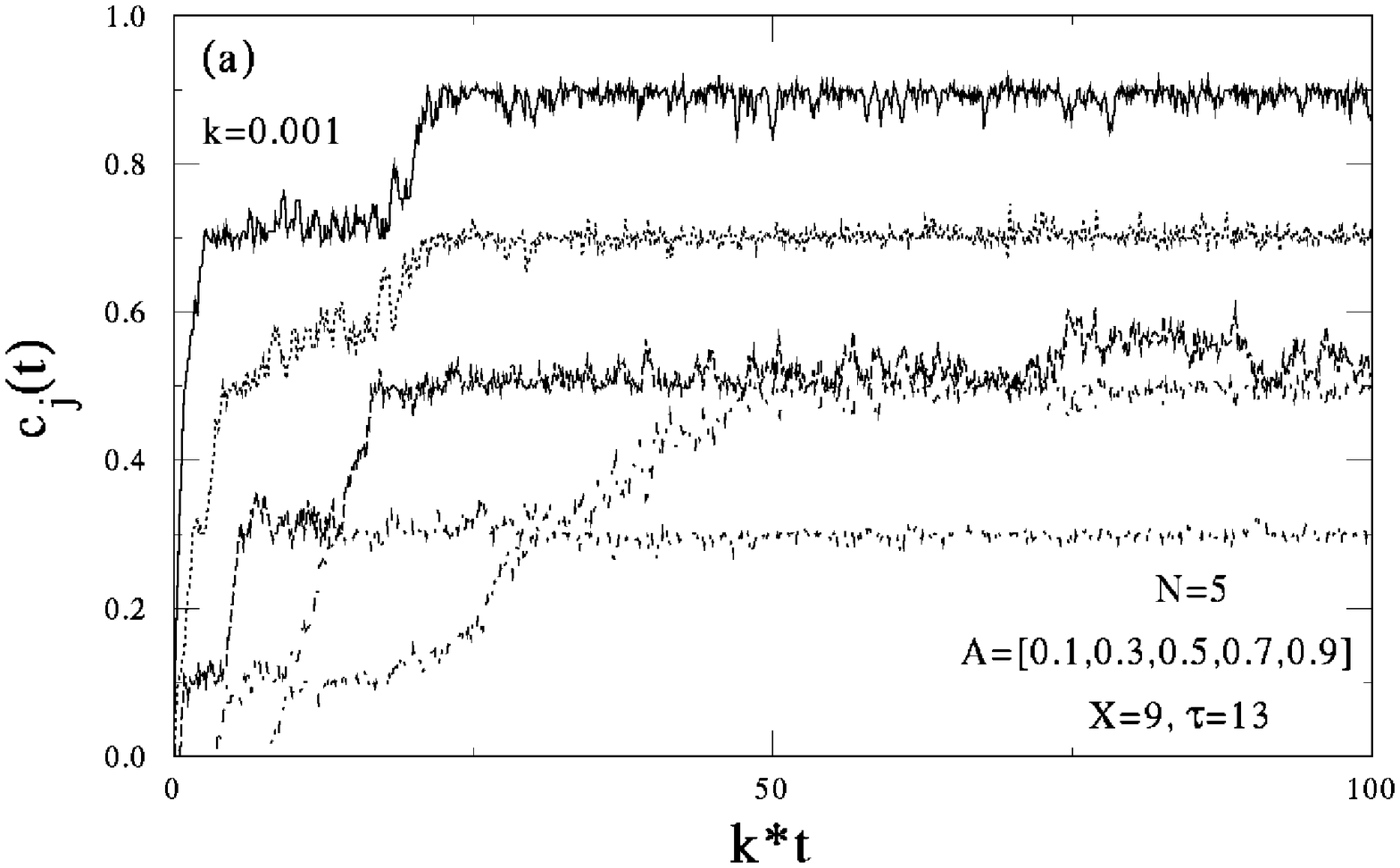}}
\vspace{-1cm}
\centerline{\epsfxsize=.8\hsize \epsfbox{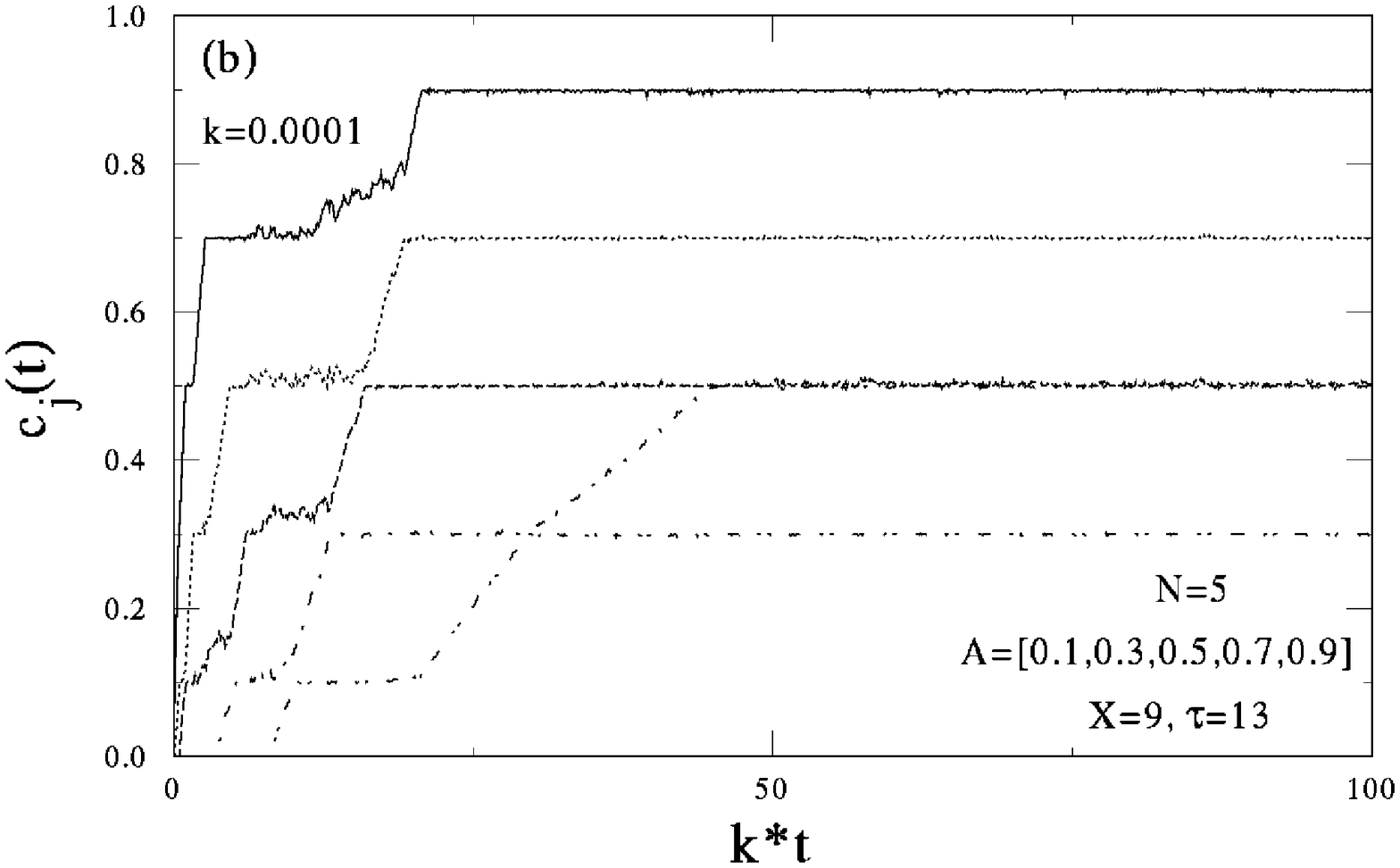}}
\vspace{.5cm}
\caption{Effect of adding noise to the system of Figure \ref{fig:nonoise},
again starting from the initial condition $x_j(t=0)=0$ for $j=1,\ldots,N$.
On each panel the curves from top to bottom
are $c_1(t)$, $\ldots$, $c_5(t)$.
Unlike the case without noise, four memory values
(0.9, 0.7, 0.5, and 0.3) all
appear to persist out to long times.
In (a), one of the curvature variables fluctuates far from memory values,
indicating that
noise can destabilize as well as stabilize memories.
In (b), the parameter $k$ has been reduced, with all other parameters
held fixed; here, the fluctuations
in the curvatures are much smaller.}
\label{fig:noise}
\end{figure}

Fig.~\ref{fig:noise}a shows the behavior of a system
identical to that of
Fig.~\ref{fig:nonoise}, except that noise has been applied
(Eq.~(\ref{eq:noisemap})), using the
parameter values $X=9$ and $\tau=13$.
Fig.~\ref{fig:noise}b
shows the time evolution in the presence of noise
for a smaller value of $k$.
Otherwise the parameters in Fig.~\ref{fig:noise}a and
Fig.~\ref{fig:noise}b are identical;
in both cases, the index $j_D$ in Eq.~(\ref{eq:noisemap})
was selected
randomly and with equal probability from the indices $1,...,N$.
Fig.~\ref{fig:noise} demonstrates that when noise is present,
more memories are stable at long times
than for the noiseless case, Fig.~\ref{fig:nonoise}.
The noise also
exhibits a destabilizing effect, as evidenced by the fluctuations in the
curvature values.
As comparison between Fig.~\ref{fig:noise}a and
Fig.~\ref{fig:noise}b demonstrates, these fluctuations become
smaller as the parameter $k$ is decreased.
Numerically we find that at times $t_0$ long enough that the
behavior appears to be stationary, the standard deviation of the
curvatures from their memory values
$\left [\frac{1}{T}\sum_{t=t_0+1}^{t _0+T} (c_j(t)-m_j)^2
\right ]^{1/2}$
is proportional to $k$ as $k \rar 0$.

%Our numerical results indicate that
%the long time behavior of the system
%does not depend on initial conditions.
Changing the parameters $X$ and $\tau$ can change
the number of different stable memories and their values.
Below we will show that the memory
values attained by each particle in
the system can be calculated
analytically by averaging
the equations of motion of the system.

\subsection{Deterministic Noise}
In the numerical simulations shown in Fig.~\ref{fig:noise}, 
the index $j_D$ was chosen randomly and
with equal probability from the indices $1,\ldots,N$.
It is useful to consider a ``deterministic'' version of noise, where
rather than selecting $j_D$ randomly,
$j_D$ is cycled systematically through the indices $1$ to $N$,
so that each index is selected exactly once during each noise cycle.
(We refer to a ``noise cycle'' as the $N\tau$ steps it takes to make
a complete cycle through the indices $1$ to $N$.)
One such choice, used for our numerics, is
to cycle through the indices in order, so that
\begin{equation}
 x_j(t+1)-x_j(t)=\floor[c_j(t)-A(t)]+ X \delta_{t (\mod~ \tau),0}~
\theta_+(j-\frac{t}{\tau}(\mod~N)+1)~,
\label{eq:detnoise}
\end{equation}
where $\theta_+(y)= 1~~{\rm if}~ y \ge 0,$ and zero otherwise.
The behavior is substantially identical for any sequence in which
each index is chosen exactly once per noise cycle.

\begin{figure}
\centerline{\epsfxsize=\hsize \epsfbox{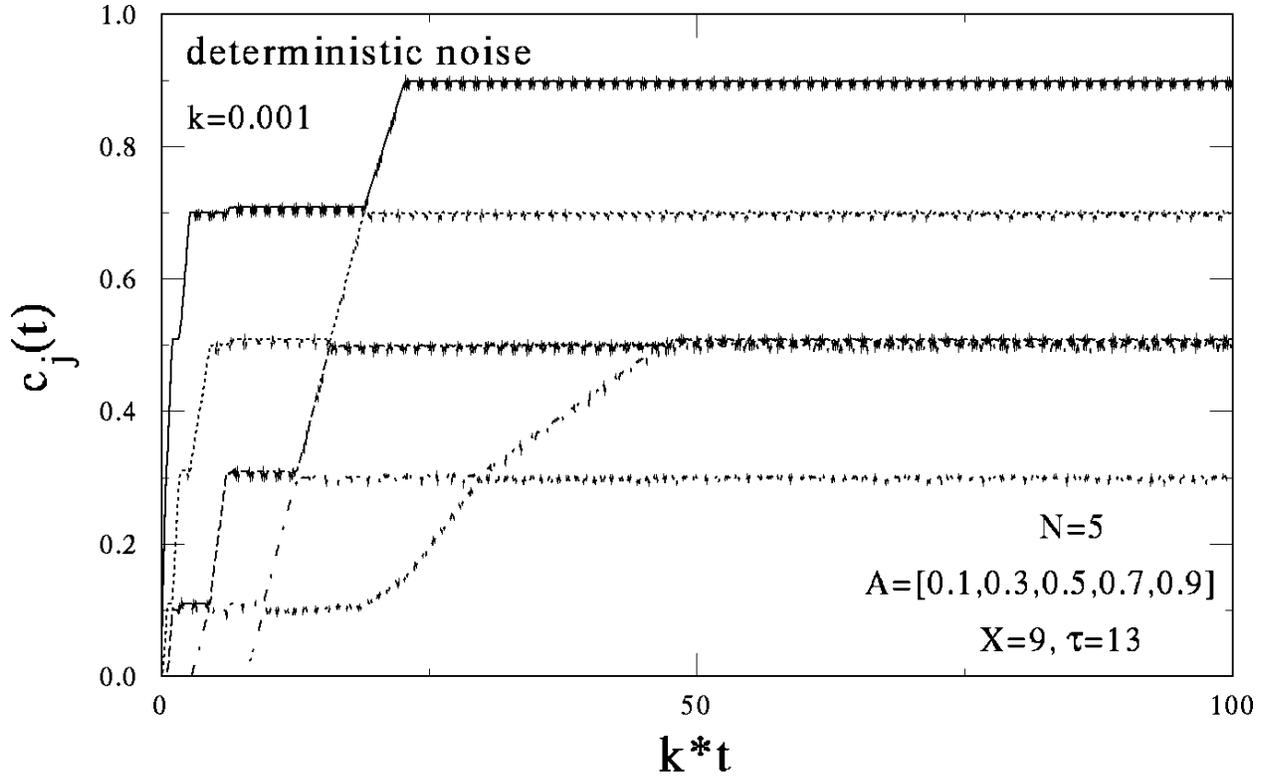}}
\vspace{.5cm}
\caption{Time evolution of the curvatures in a system of
five particles ($N=5$) in the presence of
deterministic ``noise.''
Except for order of the sequence of noise kicks, the system is
identical to that in Fig.~\ref{fig:noise}(a).
The curves from top to bottom are $c_1(t)$, $\ldots$, $c_5(t)$.
Note the similarity in the large scale features of these curves,
where deterministic noise has been applied, with
those with random noise, shown in Fig.~\protect\ref{fig:noise}.
}
\label{fig:x4_t3}
\end{figure}

\begin{figure}
\centerline{\epsfxsize=\hsize \epsfbox{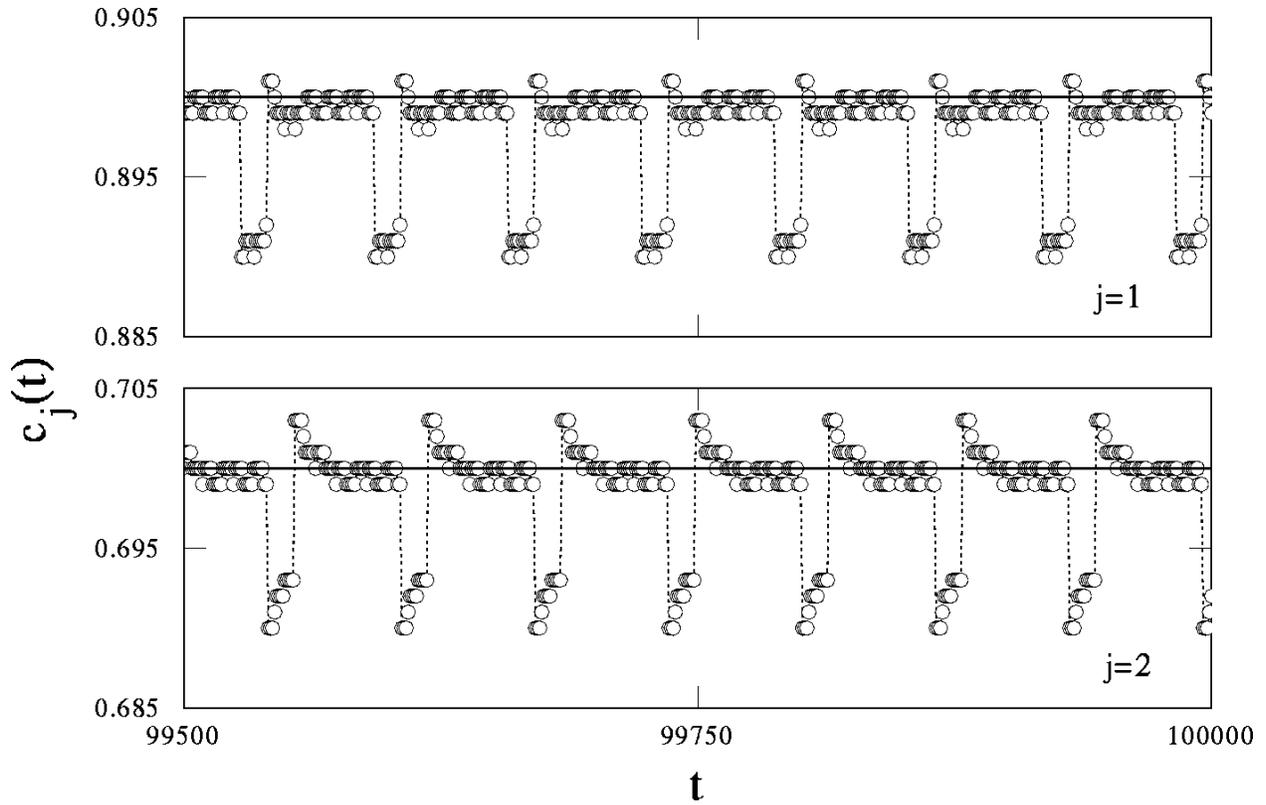}}
\vspace{2.5cm}
\caption{Plot of two curvatures
in the system of Figure~\ref{fig:x4_t3}
versus time on an expanded scale.
Parameter values are $k=0.001$, $N=5$, $A=[0.1,0.3,0.5,0.7,0.9]$,
$X=9$, $\tau=13$, with deterministic noise.
The plots show $c_1$ and $c_2$ versus time after the long-term
behavior has been reached.
The behavior is periodic; the period is $65$ time steps,
the length of one noise cycle for $N=5$ and $\tau=13$.}
\label{fig:cx4_t3}
\end{figure}

Our numerical investigations of the evolution
of Eqs.~(\ref{eq:detnoise}) over a wide range of parameters
and initial conditions indicate that
eventually the system always reaches a periodic orbit.
The period of the observed cycle is either equal to or
a divisor of $N\tau M$, where again $N$ is the number of balls,
$\tau$ is the interval between noise pulses or kicks,
and $M$ is the number of memories.
Fig. \ref{fig:x4_t3}
is a plot of the time evolution of the $c_j(t)$ 
for a five-ball system ($N=5$) with the same parameter
values as Fig.~\ref{fig:noise}, but with deterministic kicks.
The gross features of the curves are very similar, but
the fluctuations in the curvature values observed for stochastic
noise have been replaced by a regular, repeating pattern
(Fig.~\ref{fig:cx4_t3} shows an expanded view of this pattern
for two of the $c_j$'s).
The excursions during the cycles have amplitude proportional to $k$.
These regular cycles facilitate
analytic investigation of the dependence of stable memories
and their values on the parameters $X$ and $\tau$.
The number of memories remembered at long times when $k$ is small
depends systematically on the ratio $X/\tau$ and not on $X$ and
$\tau$ separately.
Figure~\ref{fig:domains} shows numerical results for the
dependence of the long-time memory values on $X/\tau$
and demonstrates the good agreement with the analytic
predictions presented in the next section.

\section{Theoretical Analysis}
\label{analytics}
In this section we show how
various aspects of the behavior of the
maps both with and without noise can be understood
analytically in the limit that $k \rightarrow 0$.
In subsection~(\ref{noisy}) we discuss the map with deterministic noise.
The observation that at long times
a periodic orbit is always reached
can be exploited to predict
the dependence of the long-time memory values on the noise parameters
$X$ and $\tau$.
A key ingredient in this analysis is the examination of the
time-averaged equations of motion of the system.
 
In subsection~(\ref{notnoisy}) we address the model without noise.
Because in this case most of the memories are transient and
therefore are no longer present when the fixed point is reached,
a modified averaging procedure must be used.
This procedure yields
insight into the transient memories and enables us to demonstrate
that a well-defined $k \rightarrow 0$ limit of this model exists.

\subsection{Long Time Behavior of The Map with Noise}
\label{noisy}
\begin{figure}[t]
\centerline{\epsfxsize=4in \epsfbox{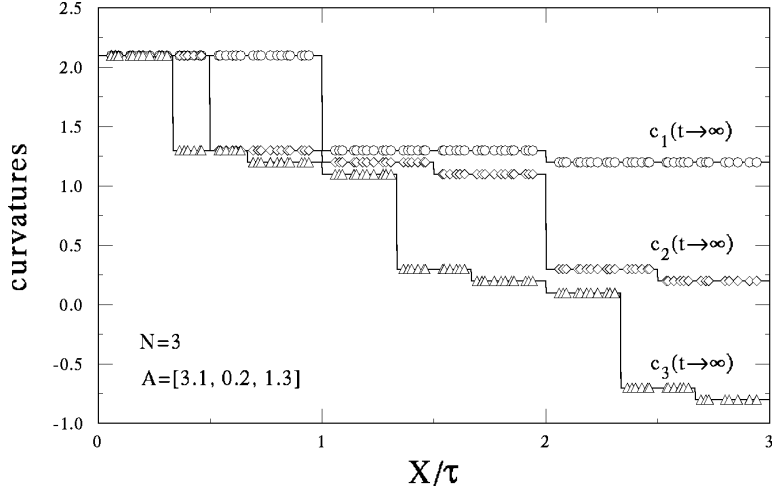}}
\vspace{.5cm}
\caption{Numerical results for the
curvature values observed at long times
as a function of the parameter $X/\tau$
for a system with $N=3$ and deterministic noise.
For these computations $k=0.0003$, but the results
are insensitive to $k$ when $k$ is small.
The solid lines are the analytical prediction for these
curvature values
using Eqs.~(\ref{eq:curv_int_part})
and (\ref{eq:domform}).
The number and value of the stable memory values
for given $X/\tau$ can be read from the graph, keeping in mind
that curvature values which differ by an integer are
on the same memory.
For example, for $X/\tau=3$, $c_1 \approx 1.3, c_2 \approx 0.3,$
and $c_3 \approx -0.7$,
so there is one stable memory for these parameter values.  Other
choices of $X/\tau$ (such as $X/\tau = 1.1$)
yield more stable memories.}
\label{fig:domains}
\end{figure}

As Fig.~\ref{fig:domains} makes evident, there is domain structure to
the dependence of the value and number of stable memories on $X/\tau$
for the map with noise, Eqs.~(\ref{eq:noisemap}).
Here we calculate analytically the structure of these domains
when $k \ll 1$ by finding the memory value
of each site as a function of the system parameter $X/\tau$.

The equation of motion for the system is
\begin{equation}
x_j(t+1)-x_j(t) = \floor[c_j(t)-A(t)] +
X\theta_+(j-j_D(t))\delta_{t (\mod~\tau),0} ~,
\label{noise_no_ave}
\end{equation}
where $\theta_+(y)$ is defined after Eq.~(\ref{eq:detnoise}).
The $j_D$'s are selected so that the probability that
$j_D(t)=n$ is $P_n$.
We examine first the case of deterministic noise and discuss
stochastic noise at the end of the subsection.

We define an averaging time $T_{ave}=NM\tau$ and
\begin{equation}
\bar{u}_j(t_0) = \frac{1}{T_{ave}}\sum_{t=t_0}^{t_0+T_{ave}-1}
\floor[c_j(t)-A(t)]~.
\label{u_jdefinition}
\end{equation}
Averaging Eq.~(\ref{noise_no_ave}) over a time $T_{ave}$ yields
\begin{equation}
\frac{1}{T_{ave}} \left (x_j(t_0+T_{ave})-x_j(t_0) \right ) =
\bar{u}_j(t_0) +
\sum_{n=1}^j P_n \frac{X}{\tau} ~.
\label{eq:noise_ave}
\end{equation}
When $t_0$ is large enough so that
$x_j(t_0+T_{ave})=x_j(t_0)$ for all $j$,
Eq.~(\ref{eq:noise_ave}) implies that
the $\bar{u}_j(t_0)$ are independent of $t_0$ (hence we
drop the argument) and must satisfy
\begin{equation}
\bar{u}_j = -\frac{ X}{\tau} \sum_{n=1}^j P_n~.
\label{u_jequation}
\end{equation}

One can also derive Eq.~(\ref{u_jequation}) directly in terms
of the curvature variables
$c_j(t) = k (x_{j+1}(t)-2x_j(t)+x_{j-1}(t))$.
Averaging Eqs.~(\ref{eq:curvnoisemap})
over a time interval $T_{ave}$ yields
\begin{eqnarray}
\frac{1}{T_{ave}}\left[c_j(t_0+T_{ave})-c_j(t_0) \right]
& = & \bar{u}_{j+1}(t_0)-2\bar{u}_j(t_0)+\bar{u}_{j-1}(t_0)
 +(P_{j+1}-P_j)\frac{X}{\tau}, ~~~ {j \ne N},\\*
& = & \bar{u}_{j+1}(t_0)-2\bar{u}_j(t_0)+\bar{u}_{j-1}(t_0)
- P_N \frac{X}{\tau},~~~~~~~~~~~~~~ {j=N}.
\label{eq:curve_noise_ave}
\end{eqnarray}
If $\frac{1}{T_{ave}}\left [c_j(t_0+MN\tau)-c_j(t_0)\right ]=0$,
as is true for a periodic orbit, one has
\begin{eqnarray}
& &\bar{u}_N - \bar{u}_{N-1} = - P_N X/\tau ~,
\nonumber
\\*
& &\bar{u}_{j+1}-\bar{u}_j + P_{j+1} {X}/{\tau} =
 \bar{u}_{j}-\bar{u}_{j-1} + P_j {X}/{\tau}~~~~(1 \le j < N)~.
\label{eq:c_ueq}
\end{eqnarray}
Eq.~(\ref{eq:c_ueq}) implies that
$\bar{u}_j - \bar{u}_{j-1} +P_jX/\tau$ is independent of $j$,
which together with the boundary conditions again yields
$\bar{u}_j = -\frac{X}{\tau}\sum_{n=1}^jP_n$.

For simplicity, assume that none of the $A_m$ are exactly
an integer\cite{Aintegerremark}
and label the values of $A_m$ such that
$0<{\rm frac}(A_1)<{\rm frac}(A_2)< \ldots < {\rm frac}(A_M)<1$.
We now show that when $k \rightarrow 0$,
every particle is almost always on a memory.
Only for a set of $P_n$ of measure zero are
some particles in the system
not on memory values as $k \rar 0$.
At long times,
the $j^{th}$ curvature is on the ${\ell_j^*}^{th}$ memory
($c_j (t \rightarrow \infty)$ obeys
${\rm frac}[c_j(t \rar \infty)] =
{\rm frac}[A_{\ell_j^*}] + {\mathcal O}(k)$),
where the memory index
$\ell_j^*$ is
\begin{eqnarray}
\ell_j^* & = & 1+\floor \left [ -M\frac{X}{\tau}\sum_{n=1}^j P_n \right ]
+ \sum_{m=1}^M \floor \left[A_m\right]
\nonumber \\*
& - & M\,  \floor \left [ - \frac{X}{\tau}\sum_{n=1}^j P_n 
+ \frac{1}{M} \sum_{m=1}^M \floor \left[ A_m\right]\right] ~.
\label{eq:domform}
\end{eqnarray}
Perhaps surprisingly, which ball is on which memory does not
depend on the memory values ${\rm frac}(A_m)$.
This analytic prediction
is completely consistent with our numerical observations;
this agreement is illustrated by the consistency of the analytic
and numerical results presented in figure \ref{fig:domains}.

We derive Eq.~(\ref{eq:domform}) by writing
$c_j(t) = {\mathcal{C}}_j + \delta c_j(t)$, where each
${\mathcal{C}}_j$ is an integer independent of $t$, and
$\delta c_j(t)$ obeys $0 < \delta c_j(t) \le 1$ for all $t$.
This decomposition can always be done if $k$ is small enough
because the maximum excursion of each $c_j$
during the averaging interval is proportional to $k$,
and every $c_j$ will turn out to be on a memory and
hence not at an integer.
We define $\delta A_m = A_m - \floor(A_m)$ and
\begin{equation}
\delta \bar{u}_{jm} = \frac{1}{N\tau}\sum_{t=t_0}^{t_0+T_{ave} -1}
\left(\floor[\delta c_j(t) - \delta A(t)]\right)\delta_{A(t),A_m} ~,
\end{equation}
and rewrite Eq.~(\ref{u_jequation}) as
\begin{equation}
-\frac{X}{\tau}\sum_{n=1}^jP_n 
=
{\mathcal{C}}_j 
- \frac{1}{M}\sum_{m=1}^M \floor(A_m)
+ \frac{1}{M} \sum_{m=1}^M \delta \bar{u}_{jm} ~.
\end{equation}
Now $\delta \bar{u}_{jm} = -1$ if $\delta c_j(t)<\delta A_m$ and
$\delta \bar{u}_{jm} = 0$ if $\delta c_j(t)>\delta A_m$ for all $t$
during the averaging interval.
Since the excursions during this interval are proportional to
$k$, they vanish
as $k \rar 0$; thus when $k$ is small enough
the $j^{th}$ ball cannot cross
more than one memory value during a cycle.
Therefore, we can write
$\sum_{m=1}^M \delta \bar{u}_{jm} = -1 + Q_j + \rho_j$,
where $Q_j$ is an integer satisfying
$0 \le Q_j \le m-1$, and the $\rho_j $ satisfy $0 < \rho_j \le 1$.
Thus we have
\begin{equation}
-\frac{X}{\tau}\sum_{n=1}^j P_n + \frac{1}{M}\sum_{m=1}^M \floor(A_m)
+1
=
{\mathcal{C}}_j
+ \frac{1}{M}(Q_j + \rho_j) ~,
\label{CQrho_equation}
\end{equation}
with ${\mathcal{C}}_j$ and $Q_j$ integers.

For simplicity we assume here that $MX/\tau\sum_{n=1}^jP_n$
is not an integer for any $j \le N$, a condition which will
ensure that $\rho_j < 1$, and hence
$\frac{1}{M}(Q_j +\rho_j) < 1$\cite{irrational_note}.
Taking the $\floor$ of both sides of Eq.~(\ref{CQrho_equation}) yields
\begin{equation}
{\mathcal{C}}_j = \floor \left [ -\frac{X}{\tau}\sum_{n=1}^jP_n
+ \frac{1}{M}\sum_{m=1}^M\floor(A_m)\right ] + 1 ~. 
\label{eq:curv_int_part}
\end{equation}
Multiplying Eq.~(\ref{CQrho_equation}) by $M$ and then taking
the $\floor$ of both sides yields
\begin{eqnarray}
Q_j & = & \floor \left [ -M\frac{X}{\tau}\sum_{n=1}^j P_n \right ]
+ \sum_{m=1}^M \floor \left[A_m\right]
\nonumber \\*
& - & M\,  \floor \left [ - \frac{X}{\tau}\sum_{n=1}^j P_n 
+ \frac{1}{M} \sum_{m=1}^M \floor \left[ A_m\right]\right]
,
\end{eqnarray}
which in turn implies
\begin{equation}
\rho_j = -M\frac{X}{\tau}\sum_{n=1}^j P_n
- \floor \left [-M \frac{X}{\tau} \sum_{n=1}^j P_n\right ]~.
\end{equation}
We see that it is consistent to assume that
$0 < \rho_j <1$ so long as $(MX/\tau)\sum_{n=1}^j P_n$ is not
exactly an integer.
Since $\rho_j$ can only be fractional if $c_j$ crosses a
memory during the averaging interval, as $k \rar 0$
each $c_j$ must be on a memory.
Since $Q_j$ determines the memory index via $\ell_j^* = Q_j+1$,
one obtains Eq.~(\ref{eq:domform}).

Finally, to demonstrate consistency of the assumption that
no particle can be on more than one memory, we must show that
the particle excursions over the averaging time are small
as $k \rar 0$.
This is easily done starting from the equation of
motion for the curvatures Eq.~(\ref{eq:curvnoisemap})
and noting that our solution for the $\bar{u}_{j}$
satisfies $\bar{u}_{j+1}-2\bar{u}_{j}+\bar{u}_{j-1}=0$.
If no memory is crossed, then the absolute value of the
difference between
the time average
$\frac{1}{T_{ave}}\sum_{t_0}^{t_0+T_{ave}} \floor[{c}_j(t)-A(t)]
\delta_{A(t),A_m}$ and the corresponding
$\floor[c_j(t)-A(t)]\delta_{A(t),A_m}$
cannot be bigger than unity.
This bound implies that until a memory is crossed,
the excursion per unit time of each of the $c$'s cannot be
bigger than $k(4+X)$.  
Since the memory values are separated by an amount of
order unity, as $k \rar 0$ the number of steps needed to
reach the nearest memory diverges as $1/k$, and
the excursion during the
averaging time cannot be greater than $k(4+X)T_{ave}$.

\subsubsection{Importance of Spatial Distribution of the Noise}
So far we have mainly discussed
the case of spatially homogeneous noise, $P_n = 1/N$ for all $n$,
and seen that if the noise causes each
spring in the chain to break with equal probability,
multiple memories can be stabilized indefinitely.
However, in our analytic work we did not assume this special form
for $P_n$, and one may ask whether the same results are obtained if,
for instance, only the first spring were broken repeatedly.

We address this issue by examining Eq.~(\ref{eq:domform}).
Note that if $P_l = 0$ for some $l$,
then particles $l-1$ and $l$ must be on the same memory value.
Thus, if only
the first spring is repeatedly broken, there will be only one memory 
observed at long times, although its value may be different than in the
noiseless case.
However, the $P_n$ need not all be equal for multiple memories
to be stable at long times.

\subsubsection{Stochastic Versus Deterministic Noise}
Now we discuss the behavior when the system is subject to stochastic
noise rather than deterministic noise.  The crucial point here is that
the equations of motion can always be averaged over some
time interval $T_{ave}$, and so long as all the $c_j$'s stay
roughly constant, there is no need for there to be a truly periodic
cycle for the procedure above to apply.
As $k \rar 0$ the memory values will be exactly the same
for random noise as for deterministic kicks with the same
time-averaged spatial distribution of events.

Not surprisingly, the excursions of the curvatures
about the memory values are larger for stochastic noise than
for sequential kicks, all other parameters being held fixed.
There are two mechanisms by which stochastic noise would enhance
the size of the excursions.
The first is that the small motions of the curvatures
about their memory values are more erratic because
the noise kicks are inhomogeneously spaced in time,
and the second is that fluctuations in the noise may temporarily
cause the system to be driven to a memory value
other than that determined by the time-averaged $X/\tau$.
Numerically we find that the excursions in systems
with stochastic noise are typically
a few times the excursions observed in the deterministic
case with the same parameters, and that their magnitude is
proportional to $k$.
These observations are evidence that the first mechanism is
dominant; the second mechanism leads to a nonlinear dependence
of the excursions on $k$ and also, since it depends on
how far each curvature is from the edge of the parameter range
in which the memory in question is stable, leads to sensitive dependence
of the excursion magnitudes on $X$ and $\tau$.

\subsection{The behavior of the map without noise as $k \rightarrow 0$}
\label{notnoisy}
In this subsection we show that a modified averaging procedure 
can be used to obtain insight into the time evolution of the
system.

Above, we used the observation that
at long times the behavior of the map with noise is periodic
in time to calculate how the memory values
depend on the system parameters.
A key step in this calculation is averaging the equations of motion
of the system over an appropriate time interval.
Here we present an averaging procedure applicable in the
limit $k \rightarrow 0$ that can be used to obtain insight
into the time evolution and not just
the long time behavior.

\begin{figure}
\centerline{\epsfxsize=\hsize \epsfbox{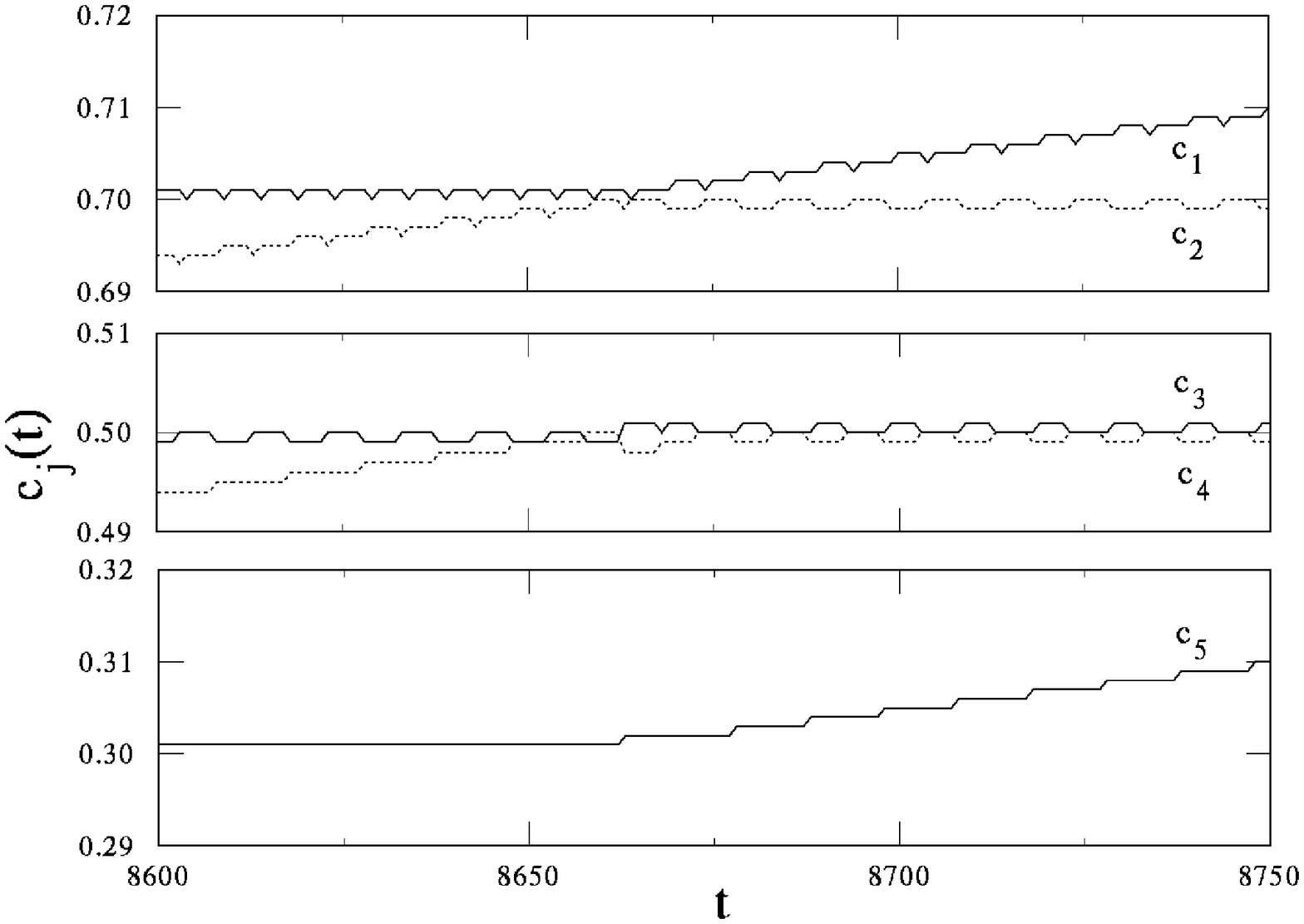}}
\vspace{.5cm}
\caption{Plot of curvatures versus time on an expanded scale
during the evolution of
the noiseless system of Figure~\ref{fig:nonoise}.
Parameter values are $k=0.001$, $N=5$, $A=[0.1,0.3,0.5,0.7,0.9]$.
The particles can be divided into two types---those whose
curvatures oscillate periodically in time about memory values
(for one case, $c_5$, the
curvature on the memory is actually time-independent until
$c_4$ hits a memory at $t \sim 8660$),
and those in transit between memory values.
}
\label{fig:nonoisedetail}
\end{figure}

Analyzing just the long-time behavior cannot yield insight
into transient memories, because in
this limit almost all the memories have been forgotten.
Therefore, the technique we used in the previous subsection
of looking only at fixed points of the equations of motion
is not so useful here.
However, Fig.~\ref{fig:nonoisedetail}, which shows the
evolution of the curvatures for the system with no noise
(the same numerical data as Fig.~\ref{fig:nonoise}
on an expanded scale), demonstrates that during the motion
two types of particles exist---sites whose curvatures are ``stuck''
on a memory values, and curvatures that are in transit between different
memory values (``drifting'').
A ``stuck'' site oscillates periodically about a memory value
until a neighbor changes its status, at which time
the stuck site can either change its oscillation about the
same memory, or can start to drift.
As $k$ is decreased,
it takes more and more time steps for the drifting sites
to get between different memories,
during which time they provide a constant environment for
their neighbors.
In contrast, the period of the cycles of the ``stuck'' sites
remains unchanged as $k \rar 0$; moreover, the amplitudes
of the excursions about the memory values are proportional to $k$.
In the limit $k \rar 0$, the drifting sites 
comprise a quasi-stationary environment for the stuck
sites, and one can average the equations of motion over
the period of the stuck sites' cycles.

For simplicity, in this subsection we consider only the
case of a single memory with $A=0$.
The generalization of the analysis to different memory
values and to multiple memories
is straightforward, and is discussed briefly in Appendix A.
We only discuss here the model in the absence of noise, but
the analysis is easily extended to the case when
noise is present, if desired.

The equations of motion for the $c_j(t)$ are
\begin{equation}
c_j(t+1) = c_j(t) + k \left ( \floor[c_{j+1}(t)]
- 2\, \floor[c_j(t)] + \floor[c_{j-1}(t)] \right )~.
\label{mdl2}
\end{equation}

First consider the behavior of a site $j$ whose two neighbors'
curvatures are both drifting between integers.
While the sites $j+1$ and $j-1$ are drifting,
the quantities $\floor[c_{j+1}(t)]$ and $\floor[c_{j-1}(t)]$
remain constant, and we can
denote their (integer) values as $I_{j+1}$ and $I_{j-1}$
and define $\eta \equiv (I_{j+1} + I_{j-1})/2$.
Eq.~(\ref{mdl2}) then becomes
\begin{equation}
c_j(t+1) - c_j(t) = 2 k \left ( \eta
-  \floor[c_j(t)] \right )~.
\label{oneball}
\end{equation}
If $\eta - \floor[c_j(t)] > 0$, then
$c_j(t)$ will increase in time until 
$\eta - \floor[c_j(t)]$ is no longer positive.
If $\eta$ is an integer, then
$c_j(t)$ will stick at $\eta$, whereas if
$\eta$ is a half-integer, then
$c_j(t)$ will undergo a period-2 cycle about the integer value
$\eta+1/2$.
If initially $\eta - \floor(c_j(t)) < 0$,
then eventually
$c_j(t)$ will stick just below $\eta+1$
if $\eta$ is an integer,
and $c_j(t)$ will oscillate in a period-2 cycle
cycle about $\eta+1/2$
if $\eta$ is a half-integer.

%*******here are the statements that would be nice to
%show analytically********
%\newline
If there are $L$ stuck sites in a row, then the
cycles of the sites become longer, but simple periodic behavior
is still observed.  We find numerically that
the motion of each site in a stuck region with $L$
sites is a cycle of length $L+1$ or shorter.  
Moreover, every time a site changes its status (for
instance, a drifting site might come within $O(k)$
of a memory value), the new cycle gets established
in a time that remains finite as $k \rar 0$.
%?and how does it depend on $L$?.
%\newline
%**********
%\newline
Therefore, as $k \rar 0$, during the periods when the
drifting sites at the boundaries of the region in question
remain between memories, we can average
Eqs.~(\ref{mdl2}) over the cycle of length $p$.
Defining $u_j = \frac{1}{p}\sum_{t=t_0+1}^{t_0+p}\floor[c_j(t)]$,
we obtain
\begin{equation}
c_j(t+p) - c_j(t) = k \left ( u_{j+1}
- 2 u_j + u_{j-1} \right )~.
\label{averaged}
\end{equation}
All the terms on the right hand side of Eq.~(\ref{averaged})
are time-independent, implying that
\begin{equation}
c_j(t_0+p) = c_j(t_0) + (kp)r_j~,
\end{equation}
with $r_j = u_{j+1} - 2 u_j + u_{j-1}$.
Moreover, we can rewrite Eq.~(\ref{mdl2}) during the averaging
interval as
\begin{equation}
c_j(t+1) = c_j(t) + kr_j + k \Delta_j(t)~,
\end{equation}
where $\Delta_j(t) = \floor[c_{j+1}(t)]-u_{j+1}
- 2(\floor[c_j(t)]-u_j) + \floor[c_{j-1}(t)]-u_{j-1}$
has zero mean and is periodic with period $p$.
Also, because $\floor(c_j)$ of a stuck site does not change
by more than $\pm 1$, we have $|\Delta_j(t)|<4$.
Therefore, we can write the complete solution of Eq.~(\ref{mdl2})
for the interval where all the stuck sites have settled into
their periodic behavior and none of the drifting sites goes
through an integer as
\beq
c_j(t_0+l) = c_j(t_0) + (kl) r_j + k \eta_j(l)~,
\label{soln}
\eeq
where $\eta_j(l)$ is periodic in time and satisfies
$|\eta_j(l)| < 4\max\{p_j\} <  4N$ for all $j$ and $l$.
The first two terms represent the piecewise linear solution
and the last term represents the cycles of amplitude of order $k$
superimposed on the piecewise linear solution.
Note that the difference between the piecewise linear part
of the solution and the complete solution
goes to zero as $k \rar 0$.

We can define a rescaled time variable $\tilde{t} = k*t$ and take
the limit of Eq.~(\ref{soln}) with $k \rar 0$, $kl$ finite,
in which the $u_j$ (and hence $r_j$) are independent of $k$.
The existence and uniqueness of this limit is demonstrated
in Appendix A.
In this limit, the solution converges to
\begin{equation}
c_j(\tilde{t_0}+\tilde{t}) = c_i(\tilde{t_0}) + r_i \tilde{t}.
\label{sol2}
\end{equation}
If a site $j$ is drifting, we set $u_j = \floor(c_j)$,
whereas if it is stuck, $u_j$ is determined by requiring
\begin{equation}
r_j = u_{j+1}-2u_j+u_{j-1} = 0~.
\label{requation}
\end{equation}
When there are $L$
stuck sites in a row (say sites $u_{j_0+1},\ldots,u_{j_0+L}$,
with $u_{j_0}$ and $u_{j_0+L+1}$ given),
then the $u_j$ in the stuck region are obtained by
solving Eq.~(\ref{requation}), yielding
\begin{equation}
u_j = u_{j_0} + \left (\frac{u_{j_0+L+1}-u_{j_0}}{L+1} \right )
(j-j_0)~.
\end{equation}
Whenever all the $u_j$ are fractional,
every site in the region must be on a memory.
The values of $u_{j_0+1}$ and $u_{j_0+L}$ enter into the
drift rates of $c_{j_0}$ and $c_{j_0+L+1}$ and hence
must be determined to obtain the time evolution of those sites.

One still needs to consider the behavior at the transitions
when the sites change between stuck and drifting.  Because
the number of steps needed to establish the new cycle structure
is finite and independent of $k$,
these transitions are instantaneous in terms of
rescaled time.
Moreover, as we demonstrate in Appendix A,
the values of the $u$'s after
each transition do not depend on the details of either
the old cycle structure or of the transition.

These considerations enable us to use the
the piecewise linear solution Eq.~(\ref{sol2}) to formulate
a $k = 0$ model.
In the $k = 0$ model, if a site $i$ is stuck, then $c_i$
is exactly an integer.
Between transitions,
\beq
\der{c_j}{t} = u_{j+1}-2 u_j + u_{j-1}~,
\eeq
where the $u_j$ are equal to the $u_j$ that are defined for a system
with non-zero $k$ by averaging between the same two transitions.
At each transition, the $c_j$ are continuous.
However, the $u_j$ change instantaneously to the new values
appropriate to the time interval after the transition but
before the next transition.

Thus we have been able to characterize
the local dynamics, and
describe the $k \rar 0$ limit of the model.
However, we have not addressed the evolution of the
large scale structure of the system.
Ref.~\cite{coppersmith97} presents numerical evidence
that all the large-scale structure of the
nonlinear equations is well-approximated by the evolution
of linear equations of motion obtained by replacing
$\floor[y]$ by $y-\half$ in Eq.~(\ref{eq:curveq}).
As discussed in Ref.~\cite{coppersmith97}, this observation
enables one to perform accurate estimates of
when memories form and when
they are forgotten as a function of system size and
model parameters.
In Appendix B we present an analytic bound on the error
in the evolution of the curvatures
made when the dynamical equations are linearized and
show that this error is bounded uniformly in time and
logarithmically in the system size.

In this subsection we have obtained the $k \rar 0$ behavior via
an explicit limiting process of the dynamics with $k\ne 0$.
In Appendix A we show that this limit is well-defined
and prescribe how to define the
$k \rar 0$ limit of the model without reference to averages of the
$k \ne 0$ dynamics.  We also sketch how to generalize the
analysis of the $k \rar 0$ limit to apply to the case of
multiple memories.

\section{Discussion}
\label{discussion}
We have investigated the behavior of a simple nonlinear dynamical
system which has the capacity to encode memories.
The deterministic system can encode many memories for a while,
but at long times forgets almost all of them.
Here we have demonstrated that there is a type of
stochastic noise which enables the system to encode
many memories permanently.
The memory stabilization arises because the 
noise has a time-average with nontrivial spatial structure;
in particular, it enables the curvature variables which
describe the local force, to have large-scale variations
even at long times.

The disappearance of the memories in the absence of noise
occurs only because the range of $c$'s
collapses at long times.
For fixed, free, or periodic boundary conditions (which seem
most appropriate to physical
realizations of balls and springs or of charge-density waves),
at long times the values of $c_0$ and $c_N$ are the same.
To stabilize many memories permanently, one must arrange things
so that $c_0 \ne c_N$ at long times.
The ``phase slip'' noise studied here is one way to do this.
In principle, another way to do this is to impose boundary
conditions which enforce $c_0 \ne c_N$, but
we do not know of a physically plausible way to do this in
the CDW system\cite{bcnote2}.

We now discuss possible consequences of our results for
experiments on CDW materials.
The experiments reported in Ref.~\cite{coppersmith97} involved
averaging over millions of applied pulses, and thus were
probably measuring
the number of memories retained in steady-state\cite{coppersmith97}.
In the experiment, the only samples which retained multiple
memories had additional
silver paint strips attached between the probe contacts.
This perturbation on the system is important, because ordinarily
one expects the phase slips to occur almost exclusively at the sample
contacts, where the strains are largest\cite{phasesliprefs}.
The silver paint in the middle of the sample
may induce spatially inhomogeneous phase slips;
our theoretical results suggest that the spatial inhomogeneity
of phase slips in a sample may be important in determining
the number of memories retained at long times.
%However, our results indicate that sometimes
%the new memory values that are stabilized
%are accidentally degenerate with those
%existing already.
Further experiments to quantify the
spatial dependence of noise in CDW materials would help determine
whether the theory is applicable to noise production
in this experimental situation.

Because the noise stabilization depends on the detailed spatial
characteristics of the noise, it will be quite sample-dependent.
On the other hand, in the absence of noise, the dependence of the
duration of the transient memories on sample size follows from
rather general arguments and should be robust\cite{coppersmith97}.
Therefore, systematic investigation of the time evolution of the
transient memory response in samples with as little noise as possible
is therefore probably the most promising avenue towards making
comparison between theory and experiment.

In this paper we have shown that it is possible to obtain a rather
complete theoretical understanding of our dynamical system in the
limit of very weak springs, $k \rightarrow 0$.
However, real CDW materials such as $NbSe_3$ tend to be described
by the model in the large-$k$ regime\cite{dicarlo94}.
Therefore, quantitative comparison between this theory and experiment
cannot be expected.
Understanding how the memory behavior evolves as $k$ is made large
and providing quantitative theoretical predictions in the regime
relevant to experiment is an important subject for future investigations.

%\section{Conclusions}
%\label{conclusions}
%We have investigated
%a dynamical system obtained from the equations of motion 
%for overdamped balls and springs
%in a periodic potential subjected to force pulses.
%This dynamical system exhibits a memory property in that the value of 
%the applied force is encoded in the configuration of the chain of balls
%and springs. In the absence of noise, the system loses all but one memory 
%in the limit of long times. It was found 
%that adding a particular type of stochastic
%noise could stabilize multiple memories for long times.
%Numerical and analytical results have been obtained to 
%characterize the memory stabilization of this type of noise.
%
\section{Acknowledgments}
S.N.C. acknowledges support by the
National Science Foundation, Award No. DMR 96-26119.
L.P.K. acknowledges support from
the Office of Naval Research,
Grant N00014-96-1-0127.
M.L.P. acknowledges support through the NSF Materials Research
Science and Engineering Center summer REU program.
This work was supported in part by the MRSEC program of the
National Science Foundation under Award Nos. DMR-9400379
and DMR-9808595.

\section*{Appendix A: The $\k \rar 0$ limit}

In Subsection~(\ref{notnoisy}) we defined the $k \rightarrow 0$
limit of the model
in terms of averages of the behavior of the model with nonzero $k$.
In this appendix, we demonstrate that this limit is well-defined and
show how to construct it without
reference to the system with nonzero $k$.
In A.I we discuss a single integer memory, while
in A.II we consider briefly the rather straightforward generalization
to the case of multiple memories.

\centerline{{\em A.I:  The $k=0$ limit}}
We recall that our original model was defined in terms
of a discrete time index $t$, and wish to introduce a
rescaled time $\tilde{t}=kt$ and consider the limit $k \rar 0$
with $\tilde{t}$ finite.
The problem we are addressing is:
Given values $c_j(\tilde{t})$ for $j = 1,2, \ldots, N$ at
some time $t$,
can the corresponding $u_j(\tilde{t})$ be generated uniquely?
If so, then because
$\der{c_j}{\tilde{t}} = u_{j+1}(\tilde{t})-2 u_j(\tilde{t})+
u_{j-1}(\tilde{t})$,
the entire time evolution is determined.

As stated, this problem is not solvable in general.
This is because, even for a model with non-zero $k$,
 the $u_j$ are not defined uniquely at a ``transition''
at which the sites go from one distribution of ``stuck''
sites or a given cycle structure to another.
As an illustration, consider the system
\beq
c_1(t+1) = c_1(t) - k ( u_0 + u_2 - 2 \floor[c_1(t)]) ~,
\eeq
where $u_0 = 0$ and $u_2 = 1$.
If initially
$c_1(0) = 1+7k/2$, we have $c_1(t) = 1 + (7-2t)k/t$,
$t \leq 4$ and $c_1(t) = 1 -(-1)^tk/2$, $t > 4$. 
On the other hand, if initially $c_1(0) = 1 - 7k/2$,
then $c_1(t) = 1 - (7-2t)k/t$, $t \leq 4$
and $c_1(t) = 1 + (-1)^tk/2$, $t > 4$.
Therefore, for $t \leq 4$, the two different initial
conditions yield $u_1(0)=1$ and $u_1(0) = 0$, respectively.
As $k \rar 0$,
both these situations correspond to the same initial
condition $c_1(0) = 1$, and therefore it is clear that
there is a transient period when $u_1(0)$ is not
defined uniquely.
Nonetheless,
%in the limit $k=0$ limit with $\tilde{t}=tk$,
since both initial conditions in the example yield
$u_1(\tilde{t}) = 1/2$ for $\tilde{t}>4k$,
in the limit $k=0$
it is consistent to define 
$u_1(\tilde{t}) = 1/2$ for all $\tilde{t}>0$.

Here we demonstrate that $u_1(\tilde{t})$ for $\tilde{t}>0$
can be defined uniquely for
all possible initial conditions
of the model with non-zero $k$ that lead to the same
initial configuration of $c$'s as $k \rar 0$.
Specifically, we show that given values $c_j(\tilde{t_0})$
for $j=1,2,\ldots,N$ there
is a unique consistent way to define $u_j(\tilde{t})$
for $\tilde{t}$ such that $\tilde{t}>\tilde{t_0}$,
valid up until the next transition.
This fact, together with the observation that the functions $c_j(t)$
are continuous, enables us to show that the $k=0$
limit of our model is well-defined.

For the $k \ne 0$ model, we have that $u_j= \floor(c_j)$
unless $c_j$ is within $O(k)$ of an integer $m$.
If we are close to a transition, we cannot define the $u_j$'s 
for all the sites that are close to integers.
However, recall that each transition takes a finite number
of steps, and hence takes up zero units of the rescaled
time $\tilde{t}=kt$ as $k \rar 0$.
The $c_j$ change in steps of $O(k)$, so the sites where $c_j$
is initially close to an integer will have a $c_j$ that is
close to the same integer after the transition.
After the transition, we have one of the following possibilities
for these sites:
\begin{enumerate}
\item The site $j$
could be stuck (more precisely, the curvature of site $j$ could
be stuck) and the value of $c_j$ will execute a cycle (possibly of
period 1) near the integer. In this case, we have
$m-1 \leq u_j \leq m$ and the site has zero average drift.
\item The site could be drifting up on average.
In this case, $c_j > m $ after the transition so that $u_j = m$.
\item The site $j$ could be drifting down on average. In this case
  $c_j < m$ after the transition so that $u_j = m-1$.
\end{enumerate}

Since we have that the average drift rate for the site $j$ after
the transition is given by
\begin{equation}
\der{c_j}{\tilde{t}} =
u_{j+1}(\tilde{t}) -2 u_j(\tilde{t}) +u_{j-1}(\tilde{t}),
\label{b1:eqofmotion}
\end{equation}
we have the following consistency conditions:
\begin{eqnarray}
  u_j(\tilde{t}) > m-1 \mbox{ implies that }
	u_{j+1}(\tilde{t})-2 u_j(\tilde{t})+u_{j-1}(\tilde{t}) \geq
  0~, \nonumber \\
  u_j(\tilde{t}) < m \mbox{ implies that }
	u_{j+1}(\tilde{t})-2 u_j(\tilde{t})+u_{j-1}(\tilde{t}) \leq 0 ~.
  \nonumber
\end{eqnarray}
These conditions are independent of $k$, and we require that the
$k = 0$ model satisfy them.  In the $k = 0$ model, a site can be stuck
only if $c_j$ is exactly an integer.  If $c_j$ is not exactly an
integer, then since $c_j(t)$ is continuous, we have
\beq
u_j(\tilde{t}_{0+}) = \floor[c_j(\tilde{t_0})]~,
\eeq
where
$\tilde{t}_{0+}=\lim_{\epsilon \rightarrow 0+} u(\tilde{t_0}+\epsilon)$.
If $c_j(\tilde{t_0}) = a$ is exactly an integer, we have that
$$
a-1 \leq u_j(\tilde{t}_{0+}) \leq a.
$$
We can combine the preceding two equations to obtain
\begin{equation}
u^{-}(c_j) \leq u_j \leq u^{+}(c_j),
\label{eq:constraint}
\end{equation}
where $u^-(x) = \lim_{\epsilon \rightarrow 0+} \floor (x - \epsilon)$
and  $u^+(x) = \lim_{\epsilon \rightarrow 0+} \floor (x + \epsilon)$.
The functions $u^+$ and $u^-$ satisfy a monotonicity
condition
\begin{equation}
u^-(a) \leq u^+(a) \leq u^-(b) \leq u^+(b)
\label{eq:monotone}
\end{equation}
for all $a < b$ (this follows from the fact that
$\floor(x)<\floor(y)$ if $x<y$).
This implies that for any given value $u$, there is
at most one value of $c$ such that it is possible for a site with
$c_j = c$ to have $u_j = u$.  This monotonicity gives the following
consistency requirement on the definition of the $u_j(\tilde{t}_{0+})$:
\begin{eqnarray}
  u_j(\tilde{t}_{0+}) > u^{-}(c_j(\tilde{t_0}))
	\mbox{ implies that }
	u_{j+1}(\tilde{t}_{0+})-
	2 u_j(\tilde{t}_{0+})+u_{j-1}(\tilde{t}_{0+})
	\geq 0 ~, \nonumber \\
  u_j(\tilde{t}_{0+}) < u^{+}(c_j(\tilde{t_0}))
	\mbox{ implies that } u_{j+1}(\tilde{t}_{0+})
	-2 u_j(\tilde{t}_{0+})+u_{j-1}(\tilde{t}_{0+}) \leq 0 ~.
\label{eq:consistency}
\end{eqnarray}
For brevity, we will henceforth suppress the time arguments; $c_j$
will represent $c_j(\tilde{t_0})$
and $u_j$ will represent $u_j(\tilde{t}_{0+})$.

We are
trying to generate the $u_j$ given the $c_j$ so that the $k = 0$ model
is well defined.
Given the $c_j$, we can take an initial condition of the form
$\tilde{c}_j = c_j + k F_j$, where the $F_j$ is a given arbitrary
bounded sequence, and
choose $k$ sufficiently small so that
$\tilde{c}_j$ is not an integer if $c_j$ is not an integer and
$|\tilde{c}_j - c_j| \leq 1/4$ for all $j$.
Then the $u_j$ obtained by following the
dynamics in a model with finite $k$ starting from this initial
condition and looking at the averages after any initial transitions
will satisfy the consistency requirements.
However, it is not clear that this procedure gives a
unique definition of $u_j$.

To show that there is only one consistent way to define $u_j$ for a
given $c_j$, we assume the opposite.
Let $u^1_j$ and $u^2_j$ be two distinct
definitions for $u_j$ that are both consistent.
Both $u^1_j$ and
$u^2_j$ satisfy the same boundary conditions,
so that $u^1_0 - u^2_0 = 0$ and $u^1_N - u^2_N = 0$.
Since $u^1 \neq u^2$, there is some index
$j^{*}$ for which $u^1_{j^*} - u^2_{j^*} \ne 0$.
Without loss of generality, we choose the
labels so that $u^1_{j^*} - u^2_{j^*} > 0$.
Since $u^1_0 - u^2_0 = 0$ and $u^1_N - u^2_N = 0$,
there must exist indices $p$ and $q$ with
$p < j^{*} < q$ such that
$u^1_p - u^2_p \leq 0$, $u^1_q - u^2_q \leq 0$,
and $u^1_j - u^2_j > 0$ for all $p < j < q$.
By Eq.~(\ref{eq:constraint}), $u^2_j \geq u^-(c_j)$,
so that $u^1_j > u^-(c_j)$ for all $p < j < q$.
Eq.~(\ref{eq:consistency}) therefore
requires that $u^1_{j+1} + u^1_{j-1} - 2 u^1_j \geq 0$
for all $p < j < q$.
Eq.~(\ref{eq:constraint}) also implies that $u^1_j \leq u^+(c_j)$,
so that $u^2_j < u^+(c_j)$ for all $p < j < q$.
Eq.~(\ref{eq:consistency}) therefore requires that
$u^2_{j+1} + u^2_{j-1} - 2 u^2_j \geq 0$ for all $p < j < q$.
Combining these two results, we have
$$
(u^1_{j+1}-u^2_{j+1}) + (u^1_{j-1} - u^2_{j-1}) - 2 (u^1_j - u^2_j)
\geq 0
$$
for all $p < j < q$.
This implies that if $u^1_j - u^2_j$ attains a
maximum on $p < j < q$, it is a constant for $p \leq j \leq q$.
Therefore,
$u^1_j - u^2_j \leq \max(u^1_p - u^2_p, u^1_q - u^2_q) = 0$
for all $p < j < q$.
This contradicts our assumption that $u^1_{j^*} - u^2_{j^*} > 0$, and
proves that there can be only one
consistent definition of $u_j$ given $c_j$.
This proves the claim from section~\ref{analytics} that
averaging over the cycles in a $k \ne 0$ model gives a consistent
$k=0$ model.

Now we present a prescription for generating the $u_j$ from the
$c_j$ without any reference to the $k \ne 0$ model.
The process consists of identifying
all sites whose $u$'s must be integers,
requiring that all the remaining $u$'s satisfy
$u_{j+1} -2 u_j +u_{j-1} = 0$, and checking to see whether
all the constraints are satisfied.  If not, then
there is at least one additional site whose $u$ is an integer,
and one such site is identified.  This process is iterated until
all the constraints are satisfied.

First consider the situation where $u_0 = u_p = 0$.
We are given a sequence $c_j$ ($0<j<p$) and hence functions
$u_i^+(c)$ and $u_i^-(c)$ that satisfy the monotonicity condition
(\ref{eq:monotone}) for each $0 < i < p$.
The dynamics of the $c_j$ are determined by Eq.~(\ref{b1:eqofmotion}),
and each $u_j$ must satisfy the constraint 
$u_j^-(c_j) \leq u_j \leq u^+_j(c_j)$.
Our earlier results generalize to this case and it follows
that there is a unique assignment of the $u_j$ that satisfy
the consistency conditions in Eq.~(\ref{eq:consistency}).
In this situation we have the following result:

\noindent{\bf Claim 1}: Let $j^{*}$ be an index where $u_j^-(c_j)$
attains a maximum for $0 < j < p$ and
$u_{j^*}^-(c_{j^{*}}) > 0$.
Then we must have $u_{j^{*}} = u_{j^{*}}^-(c_{j^{*}})$.

Proof:  Assume that this is not true. Then, we must have $u_{j^{*}} >
u_{j^{*}}^-(c_{j^{*}})$, and the consistency condition requires that
$u_{j^{*}+1} + u_{j^{*}-1} - 2 u_{j^{*}} \geq 0$.
It follows that $u_{j^*}$ is not a strict maximum
for $u_j$ for $0 \leq j \leq p$.
Let $0 < m < p$ be such that $u_m \geq u_{j^{*}}$.
Since the maximum value
for $u_j^{-}(c_j)$ was attained at $j = j^{*}$,
it follows that $u_m > u^-_m(c_m)$.
Therefore, by the preceding argument with $m$ in the
place of $j^{*}$, it follows that $u_m$ is not a strict maximum for
$u_j$. Consequently, $u_j \leq \max(u_0,u_p) = 0$ for all $0 < j < p$.
This contradicts the fact that the $u_j$ are constrained to be greater
than or equal to $u^-_j(c_j)$ and $u_{j^*}^-(c_{j^{*}}) > 0$.

A similar argument shows that if $j^{*}$ is an index where
$u_j^+(c_j)$ attains a minimum for $0 < j < p$ and
$u_{j^*}^+(c_{j^{*}}) < 0$, then
$u_{j^{*}} = u_{j^{*}}^+(c_{j^{*}})$.

If $u_j^-(c_j) \leq 0$ and $u_j^+(c_j) \geq 0$
for all $0<j<p$, the preceding result does
not give us any information.
However, in this case we can set $u_j=0$ for
$0 \leq j \leq p$.
Since this assignment satisfies the constraint and the
consistency conditions, by our earlier result, it is the
unique consistent definition for $u_j$.

Now we can solve the problem of assigning the $u_j$ given
the $c_j$ for the $k=0$ model recursively.
Assume that we know $u_p = a$ and $u_q = b$ with
$p > q$.\cite{nailedfootnote}
Let
$$
l_i =  a \frac{q-i}{q-p} + b \frac{i-p}{q-p}.
$$
For $p < i < q$ define $ \tilde{u}_i^{\pm}(c) = u_i^{\pm}(c) - l_i$
and $\tilde{u}_i = u_i - l_i$. Since $l_{j+1} + l_{j-1} - 2 l_j = 0$,
it follows that we are precisely in the situation that we
considered above.

We set $u_j = l_j$ for all $p < j < q$ and check to see if
$\tilde{u}_j^-(c_j) \leq 0$ and $\tilde{u}_j^+(c_j) \geq 0$ for all
$p < j < q$.
If not, we find an index $j^{*}$ and fix $u_{j^{*}}$
as in Claim 1 above, and repeat the procedure
for $p < j < j^*$ and $j^* < j < q$.
This determines all the $u_j$ recursively in no
more than $N$ steps.

The $u_j$ determine
the time dependence of the $c$'s via
$\der{c_j}{\tilde{t}}  = u_{j+1}-2u_j+u_{j-1}$.
The complete solution between
the transitions at $\tilde{t} = \tilde{t}_n$ and
$\tilde{t} = \tilde{t}_{n+1}$ is given by
\beq
c_j(\tilde{t}) = c_j(\tilde{t}_n)
+ r_j (\tilde{t} - \tilde{t}_n) \hspace{2pc}
\mbox{for $ \tilde{t}_n \leq \tilde{t} \leq \tilde{t}_{n+1}$} .
\eeq

\centerline{{\em A.II:  Multiple Memories}}

Now we extend our analysis of the $k \rar 0$ limit to the
case of multiple memories.
We find
\beq
\der{c_j}{\tilde{t}} = U_{j+1}-2 U_j+U_{j-1}~,
\eeq
with the $U_j$ given by
\beq
U_j = \frac{1}{p_j} \sum_{t = 0}^{p_j-1} \floor[c_j(t_0+t) - A(t_0+t)]
\eeq
if the site $j$ is stuck in a cycle of period $p_j$, and
\beq
U_j = \frac{1}{M} \sum_{t = 0}^{M-1} \floor[c_j(t_0+t) - A(t_0+t)]
\eeq
if the site $j$ is drifting,
{\em i.e.}, the fractional part of $c_j$ is not equal to the fractional
parts of any of the forcings $A(1), A(2), \ldots, A(M)$.
We define
\beq
U^- = \lim_{\ep \rightarrow 0+} \frac{1}{M}
\sum_{m = 0}^{M-1} \floor[c_j -\ep - A(m)]
\eeq
and
\beq
U^+ = \lim_{\ep \rightarrow 0+} \frac{1}{M}
\sum_{m = 0}^{M-1} \floor[c_j +\ep - A(m)] ~.
\eeq

Then we can assign the stuck sites and the drifting sites by
the same procedure as for the single memory except that we
replace $u^-$ by $U^-$ and $u^+$ by $U^+$.

\section*{Appendix B:  The linearized map}
In this appendix, we address the large-scale dynamics of the
system by examining a linearized equation obtained by
approximating the $\floor$ function in Eq.~(\ref{eq:curveq})
with $z-1/2$, yielding the linearized map:
\begin{equation}
c_j(t+1)-c_j(t) = k\left [c_{j+1}(t) - 2c_j(t) + c_{j-1}(t) \right ]~.
\label{eq:linearized}
\end{equation}
Although this linearized map contains no information about
the memory formation, it captures
accurately the behavior of the system at large scales.
Ref.~\cite{coppersmith97} presented numerical evidence for this
observation, and showed that it enables one
to obtain analytic estimates on the dependence of the
memory formation and forgetting processes on system size and
model parameters.

This appendix has two subsections.  In the first, we
present an analytic bound on the difference between the
configurations generated by the linear and nonlinear
equations starting from the same initial conditions.
This bound on the difference grows
logarithmically with system size, which is very slowly indeed.
Therefore, although the memories are absent in the linearized equation
(indeed, the $A(t)$ drop out entirely), the linearized
equation yields a very accurate description of the system's
behavior on large scales.

The second subsection discusses the effect of the noise on
the linearized map.  We demonstrate that the difference between
the configurations yielded by the nonlinear and the linear
equations differ by no more than unity, and that this bound
is saturated in some situations.

\subsection{Analytic bounds on behavior on large scales for
the map without noise}
In this subsection we present an analytic bound on
the error in the curvatures that is made when one approximates
the full nonlinear Eq.~(\ref{eq:curveq}) with the linearized version
Eq.~(\ref{eq:linearized}).
As discussed in Ref.~\cite{coppersmith97}, numerically we
observe that the error in the curvatures made by approximating
the nonlinear equation with the linearized one is of order
unity for all system sizes, boundary conditions, and
initial conditions.
The analytic bound presented here, valid for the $k \rar 0$
limit of the model, demonstrates that
the difference between the
configurations of the two equations is bounded by an amount
independent of time and which increases only logarithmically
with the size of the system.
This result provides further evidence
that the long wavelength behavior of the nonlinear equations
(though not the memory formation itself)
can be estimately accurately using the linearized equations.

We proceed by writing the equation of motion for the nonlinear
system, Eq.~(\ref{eq:curveq}), in the limit $k = 0$ as
\begin{equation}
\frac{dc_j(t)}{dt} = c_{j+1}(t) - 2c_j(t) + c_{j-1}(t)
- (\delta_{j+1}(t)- 2 \delta_j(t) + \delta_{j-1}(t))~,
\label{deltaequation}
\end{equation}
where $\delta_j(t) \equiv {\rm frac}(c_j - A(t))-1/2$.
The definition of frac, the fractional part function, implies that
$-1/2 \le \delta_j(t) < 1/2$ for all $j$ and $t$.
For brevity, here we drop the tilde and use $t$ to denote
a continuous time variable.
We compare the solution to Eq.~(\ref{deltaequation}) to
that of the (linear) equation where $\delta_j(t)=0$
for all $j$ and $t$, starting from the same initial conditions.
We denote the solution to the nonlinear equation $c_j(t)$
and the solution to the linearized equation as $l_j(t)$.

We define
\begin{eqnarray}
A_q(t) = \frac{1}{\sqrt{N}}\sum_j e^{iqj} c_j(t) , \\
B_q(t) = \frac{1}{\sqrt{N}}\sum_j e^{iqj} \delta_j(t) ~,
\end{eqnarray}
and Fourier transform Eq.~(\ref{deltaequation}), obtaining
\begin{equation}
\frac{dA_q(t)}{dt} = -\omega_q(A_q(t)-B_q(t))~,
\label{ftequation}
\end{equation}
with $\omega_q \equiv 2(1-\cos q)$.
This equation has the solution\cite{r_note}
\begin{equation}
A_q(t) = e^{-\omega_q t}A_q(t=0)
+ \omega_q e^{-\omega_q t} \int_0^t d\tprime e^{\omega_q \tprime}
B_q(\tprime)~.
\label{A_qequation}
\end{equation}
Note that the first term on the right
hand side of Eq.~(\ref{A_qequation}) is just $l_j(t)$, the
solution to linearized equation with $\delta = 0$.
Therefore, if we define the deviations from the linearized
solutions
\begin{equation}
\Delta c_j(t) = c_j(t)  - l_j(t) ~
\end{equation}
and
\begin{equation}
\Delta A_q(t) = A_q(t)  - 
 \frac{1}{\sqrt{N}}\sum_j e^{iqj} l_j(t) ,
\end{equation}
then
\begin{equation}
\Delta A_q(t) = \omega_q e^{-\omega_q t}
\int_0^t d\tprime e^{\omega_q \tprime}B_q(\tprime)~.
\end{equation}
Fourier transforming, we obtain
\begin{eqnarray}
\Delta c_j(t) & = & \frac{1}{N} \sum_{\jprime} \sum_q
e^{iq(\jprime - j)} \omega_q e^{-\omega_q t}
\int_0^t d\tprime e^{\omega_q \tprime} \delta_\jprime (\tprime)~\\
& = & \sum_\jprime \int_0^t d\tprime \delta_\jprime (\tprime)
\frac{d}{d\tprime} G_{j - \jprime}(t - \tprime) ~,
\label{deltac_jequation}
\end{eqnarray}
where
\begin{equation}
G_{j - \jprime}(t - \tprime) = \frac{1}{N}\sum_q e^{iq(\jprime - j)}
e^{\omega_q (\tprime - t)} ~
\label{greenfunction}
\end{equation}
is the Green's function specifying the response
at site $j$ and time $t$ to a disturbance at site
$\jprime$ and time $\tprime$\cite{kadanoff62,doniach74}.

To proceed further, we investigate the
the Green's function, Eq.~(\ref{greenfunction}).
Up to now our manipulations have been exact for any length
chain, but now we specialize to the case of long chains,
for which the sum over $q$ can be replaced by an integral,
yielding (Ref.~\cite{abramowitz72}, 9.6.19):
\begin{equation}
G_{j - \jprime}(\tau) = I_{\jprime-j}(2\tau) e^{-2\tau}~,
\end{equation}
where $I_\nu(x)$ is the modified Bessel function of the
first kind of order $\nu$.
If $\jprime =j$, then $G_{j - \jprime}(\tau)$ monotonically
decreases from $1$
to $0$ as $\tau$ goes from $0$ to $\infty$, while for
$\jprime \ne j$, $G_{j - \jprime}(\tau)$ has a single maximum
as a function of $\tau$;
it rises from zero to a maximum value and then decreases
back to zero at large $\tau$.
At large distances and long times, the contribution of large $q$'s
is suppressed exponentially, so that it is very accurate to
approximate $\omega_q$ with its small $q$ limit,
$\omega_q \approx q^2$, yielding the
Green's function\cite{chandrasekhar43}:
\begin{equation}
G_{j - \jprime}(t-\tprime) \cong \frac{1}{\sqrt{4\pi (t-\tprime)}}
\exp \left [-\frac{(\jprime - j)^2}{4 (t-\tprime)}\right ]~.
\end{equation}
This function has its maximum when $(t-\tprime)=j^2/2$, with the
value $G^*_{\jprime-j}=(\sqrt{2\pi e} |j-\jprime |)^{-1}$.

The simple behavior of the $G$'s, together with the bounds
$-1/2 \le \delta_\jprime(\tprime) < 1/2$ for all $\jprime$
and $\tprime$, can be used
to bound $|\Delta c_j|$.
The absolute value of the right hand side
of Eq.~(\ref{deltac_jequation}) is maximized if $\delta=1/2$
whenever the time derivative of $G$ is positive, and
$\delta=-1/2$ whenever the time derivative of $G$ is negative.
Thus, one obtains the bound for long chains:
\begin{eqnarray}
|\Delta c_j(t)| & \le &
 1+ \sum_{\jprime \ne j} G^*_{\jprime-j} \nonumber \\
& \approx & 1+
\sum_{\jprime \ne j} (\sqrt{2\pi e} |j-\jprime |)^{-1} \nonumber \\
& \propto & \ln(N) ~,
\end{eqnarray}
where, again, $N$ is the length of the chain.
In more dimensions, a similar calculation yields the result
that the bound grows logarithmically with the linear dimension
of the system.
Thus, the linearized map deviates from the exact solution by
an amount that is bounded at all times by an amount that
grows very slowly with system size.
The deviations observed numerically are smaller than this
bound; this is not surprising because the bound is obtained
for a particular choice of correlated $\delta$'s, which is
unlikely to be generated by the dynamics.

\subsection{Linearized Map With Noise}
A bit more insight into the linearized equation can be obtained
by investigating the long-time behavior of the
linearized map with noise added for the ``nailed" boundary condition.
We show that the
difference between the curvature values in the linearized solution and
the nonlinear
solution is bounded above by an amount of order unity.

We start with Eq.~(\ref{u_jequation}) together with
Eq.~(\ref{u_jdefinition}), yielding
\begin{equation}
\frac{1}{T_{ave}}\sum_{t=t_0}^{t_0+T_{ave}-1} \floor[c_j(t)-A(t)]
= -\frac{X}{\tau}\sum_{n=1}^j P_n~.
\end{equation}
We linearize this equation by replacing $\floor(z) \rar z-\half$
and obtain
\begin{equation}
\overline{c^{linear}_j}=
{-X \over \tau}\sum_{n=1}^j P_n +\overline{A} + \half ~.
\end{equation}

We can compare this result with that for the
nonlinear equations.  To do this, we can use the
bound ${\mathcal{C}}_j < c^{nonlinear}_j(t) \le {\mathcal{C}}_j +1$
(recall $c_j(t) = {\mathcal{C}} + \delta c_j(t)$, with
$0 < \delta c_j(t) \le 1$),
and cast Eq.~(\ref{CQrho_equation}) as the inequality
\begin{equation}
c^{nonlinear}_j(t) \le
-\frac{X}{\tau}\sum_{n=1}^j P_n + \frac{1}{M}\sum_{m=1}^M\floor[A_m]+1
< c^{nonlinear}_j(t) + 1 ~.
\end{equation}
Using the inequalities $x-1 < \floor[x] \le x$, we find
\begin{eqnarray}
c^{nonlinear}_j(t) < \overline{c^{linear}_j} + \half~,
\nonumber \\*
c^{nonlinear}_j(t) > \overline{c^{linear}_j} - \frac{3}{2}~.
\end{eqnarray}
The difference between $\overline{c^{linear}_j}$
and $\overline{c^{nonlinear}_j}$ is thus bounded by an amount
of order unity even as the system size $N \rar \infty$.
Thus, though once again the linearized model does not yield
information about the memory values exhibited by the
system, it does provide an accurate description of the
large scale variations of the configuration.

%%%%%%%%%%%%%%%%%%%%%%%%%%%%%%%%%%%%%%%%%%%%%%%%%%%%%%%%%%%%%%%%%%%%%%%
% the figures follow here


\begin{thebibliography}{99}
\bibitem{ballfootnote}
	This relationship is easiest to understand in the limit that the
	potential is turned off during each force pulse.
	During each pulse, each particle translates by an amount
	proportional to the total force on it
	(applied force plus spring force).
	causing each particle to fall into the nearest well
	bottom, at the closest integer.
\bibitem{signfootnote}
	The signs here are chosen so that positive $X$ corresponds to
	the balls being moved closer to the nail (each noise event
	causing a reduction in strain).
\bibitem{bc_note}  The boundary conditions for the $c$'s is obtained
	most easily by writing $x_j(t+1)-x_j(t) = \floor(c_j(t)-A(t))$.
	The requirement $x_0(t)=0$ for all $t$ can be satisfied by
	setting $c_0(t)=A(t)$ for all $t$.
\bibitem{springfootnote} Let $u_j(t)$
	be the position of the $j^{th}$ particle and $a_j(t)$ be the
	unstretched length of the spring between particles $j$ and $j-1$
	at time $t$, so that the spring force exerted by the $j^{th}$
	spring on particle $j$ at time $t$ is
	$-k(u_j(t) - u_{j-1}(t) + a_j(t))$.
	(The sign follows because
	the force pulses push the particles backward, so that
	the $u_j$'s are negative and the curvatures
	(spring forces) are positive.)
	The $u_j(t)$ must satisfy:
	\begin{eqnarray}
	& ~ &u_j(t+1) - u_j(t) = \hfil \nonumber \\*
	& ~ & \floor \left [ k \left [
	(u_{j+1}(t)-u_j(t)+a_{j+1}(t)) -
	(u_j(t) - u_{j-1}(t)+a_j(t)) \right ]
	- A(t) \right ] \ .
	\label{eq:uequation}
	\end{eqnarray}
	Defining $x_j(t) \equiv u_j(t) - d_j(t)$, with
	$d_j(t) = \sum_{l=0}^j a_j(t)$,
	we find
	\begin{eqnarray}
	& & x_{j+1}(t+1) - x_j(t) + [d_j(t+1)-d_j(t)] =
	\hfil \nonumber \\*
	& & ~~~
	\floor \left [ k(x_{j+1}(t)-2x_j(t)+x_{j-1}(t)) - A(t) \right ] ~.
	\end{eqnarray}
	If at time $t$, the
	unstretched length of the ${j_D}^{th}$ spring, $a_{j_D}(t)$,
	is increased by $X$,
	we obtain Eq.~(\ref{eq:noisemap}).
\bibitem{Aintegerremark}
It is straightforward to generalize the analysis presented here
to include the case where an $A_m$ is an integer, but at the
cost of a more cumbersome notation.
\bibitem{bcnote2} To make $c_0$ and $c_N$ different at long times,
	one needs to apply forces at the ends whose magnitudes
	or durations
	are accelerating at different rates.
	Using force pulses which are inhomogeneous in space but
	independent of time does not lead to multiple memories,
	as can be seen from the results with no noise and
	the nailed boundary condition.
\bibitem{r_note}  One way to derive this solution is to define
	$R_q(t) = \exp[\omega_q t] A_q(t)$.
	Using Eq.~(\ref{ftequation}), one finds that the $R_q(t)$ obey
	\begin{equation}
	\frac{dR_q(t)}{dt} = \omega_q e^{\omega_q t} B_q(t)~,
	\end{equation}
	which has the solution
	\begin{equation}
	R_q(t) = R_q(t=0) + \omega_q \int_0^t d\tprime
	e^{\omega_q \tprime}B_q(\tprime)~.
	\end{equation}
\bibitem{nailedfootnote}  This is clearly possible when the boundary
	conditions specify $u_0$ and $u_N$.  Our nailed boundary
	condition of one free and one fixed end is equivalent to
	fixing $u_0 = u_{2N+1} = 0$, with $u_{2N+1-j}=u_j$.
\end{thebibliography}
\end{document}